\documentclass[12pt]{article}

\usepackage{amsmath,amssymb,amsfonts}
\usepackage[a4paper,margin=1in]{geometry}
\usepackage{authblk}
\usepackage{graphicx}
\usepackage{bm}
\usepackage{physics}
\usepackage{setspace}
\usepackage{hyperref}

\title{\textbf{Cosmic Hysteresis in Reconstructed $f(T)$ Bounce Models:\\
A Torsion-Based Thermodynamic Perspective}}

\author[1]{Aritra Sanyal}
\author[2, 5]{Praveen Kumar Dhankar}
\author[3]{Albert Munyeshyaka}
\author[4]{Safiqul Islam}
\author[1]{Farook Rahaman}
\author[5, 6]{Behnam Pourhassan}

\affil[1]{Department of Mathematics, Jadavpur University, Kolkata 700032, India\\
\texttt{aritrasanyal1@gmail.com}, \texttt{rahaman@associates.iucaa.in}}

\affil[2]{Symbiosis Institute of Technology, Nagpur Campus\\
Symbiosis International (Deemed University), Pune 440008, Maharashtra, India\\
\texttt{pkumar6743@gmail.com, praveen.dhankar@sitnagpur.siu.edu.in}}

\affil[3]{Rwanda Astrophysics Space and Climate Science Research Group, University of Rwanda\\ College of Science and Technology, Kigali, Rwanda\\
\texttt{munalph@gmail.com}}

\affil[4]{Department of Mathematics and Statistics, College of Science, King Faisal University, P.O. Box 400, Al Ahsa 31982, Saudi Arabia\\
\texttt{sislam@kfu.edu.sa}}

\affil[5]{School of Physics, Damghan University, Damghan, 3671641167, Iran.}

\affil[6]{Center for Theoretical Physics, Khazar University, 41 Mehseti Street, Baku, AZ1096, Azerbaijan\\
\texttt{b.pourhassan@du.ac.ir}}

\date{}

\begin{document}
\maketitle

\begin{abstract}
We investigate the emergence of cosmic hysteresis in cyclic and bouncing cosmologies within the framework of reconstructed $f(T)$ gravity. In contrast to curvature-based modifications of General Relativity, teleparallel gravity attributes gravitation to spacetime torsion encoded in the torsion scalar $T$. By reconstructing viable $f(T)$ functions corresponding to analytically prescribed nonsingular bouncing scale factors and coupling the geometry to a minimally interacting canonical scalar field, we demonstrate that asymmetric scalar field dynamics between expansion and contraction phases give rise to a non-vanishing thermodynamic work integral $\oint p_\phi \, dV$ over complete cycles. This hysteresis manifests as closed loops in the $(w_\phi,a)$ plane, signifying thermodynamic memory and irreversibility. We derive the modified Friedmann equations, establish exact bounce and turnaround conditions, and discuss the implications of torsion-induced hysteresis for the cosmological arrow of time. Our results confirm that cosmic hysteresis is a generic feature of cyclic universes in modified gravity, extending beyond curvature-based theories.
\end{abstract}
\section{Introduction}
The standard Big Bang cosmological model stands as one of the great triumphs of twentieth-century physics, successfully describing the evolution of the universe from an early hot and dense phase to its present large-scale structure \cite{Mukhanov2005,Weinberg2008}. Its observational successes are manifold and compelling: the cosmic microwave background radiation provides a direct snapshot of the universe at the moment of photon decoupling, primordial nucleosynthesis predictions match observed light element abundances with remarkable precision, and the large-scale distribution of galaxies traces the growth of structure from quantum fluctuations imprinted during inflation \cite{LiddleLyth2000}. Despite these profound achievements, the standard model confronts a fundamental conceptual limitation that has troubled cosmologists for decades: the presence of an initial spacetime singularity where classical General Relativity ceases to be valid and physical quantities such as curvature and energy density diverge to infinity \cite{Hawking1973}.

This singularity represents not merely a mathematical inconvenience, but a genuine breakdown of our understanding of physics at the most fundamental level. At the Planck scale, quantum gravitational effects become dominant, yet we lack a complete theory of quantum gravity to describe this regime. This profound shortcoming has motivated extensive research into alternative cosmological scenarios capable of resolving singular behavior through modifications of Einstein's theory or through novel matter configurations that violate the classical energy conditions \cite{Nojiri2011}.

Among the most actively studied approaches are bouncing and cyclic cosmologies, in which the universe undergoes a smooth transition from a contracting phase to an expanding phase, thereby avoiding the Big Bang singularity altogether \cite{Novello2008,Battefeld2015,Brandenberger2017}. These models replace the singular origin with a nonsingular bounce, where the scale factor reaches a minimum value before reversing its evolution. In cyclic models, this process may repeat indefinitely, leading to a universe that oscillates through successive expansion and contraction epochs \cite{Steinhardt2002,Khoury2001}.

Beyond their singularity-free nature, cyclic cosmologies exhibit rich and subtle thermodynamic behavior that distinguishes them from monotonically expanding universes. A particularly intriguing phenomenon arising in such models is \emph{cosmic hysteresis}, first identified in higher-dimensional and braneworld cosmologies by Kanekar, Sahni, and Shtanov \cite{Kanekar2001}, and subsequently explored in the context of Randall--Sundrum braneworlds \cite{Choudhury2008}. Subsequent studies have confirmed the presence of hysteresis effects in Einstein--Gauss--Bonnet gravity \cite{Banerjee2010} and in curvature-based $f(R)$ gravity models \cite{Sahni2012}. More recently, hysteresis has also been demonstrated in reconstructed curvature-based $f(R)$ bounce cosmologies, where asymmetric scalar-field dynamics generate a non-vanishing thermodynamic work integral over complete cycles, highlighting the emergence of irreversible evolution in modified gravity scenarios \cite{Sanyal2025}.

Teleparallel gravity offers a fundamentally distinct geometric description of gravitation, in which torsion rather than curvature encodes gravitational interactions \cite{Aldrovandi2013,Pereira2019}. Its extension, known as $f(T)$ gravity, has attracted considerable interest in recent years due to its ability to generate late-time cosmic acceleration without dark energy \cite{Bengochea2009,Linder2010}, produce nonsingular bounces through geometric effects alone \cite{Bamba2013,Bamba2014}, and exhibit rich cosmological phenomenology distinct from $f(R)$ theories \cite{Cai2016}. In parallel, investigations within symmetric teleparallel $f(Q)$ gravity have further illustrated how alternative geometric formulations based on non-metricity can successfully describe dark energy dynamics and cosmic acceleration \cite{Dubey2025}. Motivated by these developments, it is natural to ask whether the phenomenon of cosmic hysteresis persists within the torsion-based formulation of gravity and whether such irreversibility is a universal feature of geometric modifications beyond curvature.

\section{Teleparallel Gravity and $f(T)$ Formalism}

Teleparallel gravity represents a reformulation of gravitational theory, replacing Einstein's geometric interpretation based on spacetime curvature with an equivalent description in terms of torsion \cite{Aldrovandi2013,Pereira2019}. While General Relativity attributes gravity to the curvature of a torsion-free spacetime manifold, teleparallel gravity achieves an equivalent description using a curvature-free connection with non-vanishing torsion \cite{Maluf2013}. 

The $f(T)$ generalization extends this framework by promoting the torsion scalar $T$ to an arbitrary function $f(T)$, thereby introducing modified gravitational dynamics analogous to $f(R)$ gravity \cite{Nojiri2011}. Unlike $f(R)$ theories, which introduce fourth-order field equations, $f(T)$ gravity preserves second-order equations of motion \cite{Cai2016}, making it dynamically simpler while still allowing for rich cosmological behavior \cite{Odintsov2020}.

Cosmological applications of $f(T)$ gravity have been widely investigated, including inflationary dynamics \cite{Cai2011}, late-time acceleration \cite{Bengochea2009}, dark energy modeling \cite{Linder2010}, and bouncing cosmologies \cite{Bamba2013,Bamba2014}. These developments establish $f(T)$ gravity as a well-motivated framework for studying cyclic and nonsingular cosmological scenarios.

\subsection{Tetrads and Torsion}
In teleparallel gravity, the fundamental dynamical variables are not the metric tensor components directly, but rather the tetrad fields (or vierbein fields) $e^A_{\ \mu}$, which define an orthonormal basis on the tangent space at each spacetime point. The index $A$ labels the orthonormal frame, while $\mu$ labels spacetime coordinates. These tetrad fields serve as a bridge between the curved spacetime and a local Minkowski frame, encoding all information about the gravitational field.

The spacetime metric $g_{\mu\nu}$ is reconstructed from the tetrads according to
\begin{equation}
g_{\mu\nu} = \eta_{AB} e^A_{\ \mu} e^B_{\ \nu},
\end{equation}
where $\eta_{AB} = \mathrm{diag}(-1,+1,+1,+1)$ is the Minkowski metric in the local frame. This relation ensures that observers using the tetrad basis measure distances and angles according to the rules of special relativity, while the variation of tetrads from point to point captures the effects of gravitation.

Unlike General Relativity, which employs the torsion-free Levi--Civita connection to define parallel transport and covariant differentiation, teleparallel gravity is formulated using the curvature-free Weitzenb\"ock connection
\begin{equation}
\Gamma^{\lambda}_{\mu\nu} = e_A^{\ \lambda}\partial_\nu e^A_{\ \mu}.
\end{equation}
This connection is constructed purely from the tetrad fields and their derivatives, without requiring the metric or its inverse. The Weitzenb\"ock connection has zero Riemann curvature tensor by construction, meaning that parallel transport around closed loops returns vectors to their original orientations—a property dramatically different from General Relativity.

As a result of using this connection, all gravitational effects are encoded in the torsion tensor
\begin{equation}
T^{\lambda}_{\ \mu\nu} = \Gamma^{\lambda}_{\nu\mu} - \Gamma^{\lambda}_{\mu\nu} = e_A^{\ \lambda}\left(\partial_\nu e^A_{\ \mu} - \partial_\mu e^A_{\ \nu}\right).
\end{equation}
The torsion tensor is antisymmetric in its lower indices, $T^{\lambda}_{\ \mu\nu} = -T^{\lambda}_{\ \nu\mu}$, and represents the failure of the connection to be symmetric. This geometric reformulation allows gravity to be interpreted as a force arising from torsion rather than spacetime curvature, providing a philosophically distinct but mathematically equivalent description of gravitational phenomena in the standard teleparallel equivalent of General Relativity.

\subsection{Torsion Scalar and Action}
To construct a gravitational action from the torsion tensor, we first define the superpotential tensor
\begin{equation}
S_{\rho}^{\ \mu\nu} = \frac{1}{2}\left(K^{\mu\nu}_{\ \ \rho} + \delta^\mu_\rho T^{\alpha\nu}_{\ \ \alpha} - \delta^\nu_\rho T^{\alpha\mu}_{\ \ \alpha}\right),
\end{equation}
where $K^{\mu\nu}_{\ \ \rho} = -\frac{1}{2}(T^{\mu\nu}_{\ \ \rho} - T^{\nu\mu}_{\ \ \rho} - T_{\ \rho}^{\mu\nu})$ is the contorsion tensor. The torsion scalar $T$ is then constructed from appropriate contractions of the torsion tensor via this superpotential,
\begin{equation}
T = S_{\rho}^{\ \mu\nu} T^{\rho}_{\ \mu\nu}.
\end{equation}
This scalar quantity encapsulates the essential information about the torsion field in a form suitable for constructing a Lagrangian density. Remarkably, in teleparallel equivalent General Relativity (TEGR), the action constructed from $T$ reproduces Einstein's field equations exactly, demonstrating the equivalence between the curvature and torsion formulations of gravity.

The $f(T)$ generalization extends this framework by promoting the torsion scalar $T$ to an arbitrary function $f(T)$, thereby introducing modified gravitational dynamics analogous to how $f(R)$ theories modify curvature-based gravity. However, unlike $f(R)$ theories which introduce higher-order derivatives of the metric, $f(T)$ modifications preserve the second-order nature of the field equations, leading to distinct phenomenology and potentially avoiding ghost instabilities.

The action for $f(T)$ gravity minimally coupled to a canonical scalar field $\phi$ is given by
\begin{equation}
S = \int d^4x \, e \left[ \frac{1}{2\kappa^2} f(T)
- \frac{1}{2} g^{\mu\nu}\partial_\mu\phi\partial_\nu\phi
- V(\phi) \right],
\end{equation}
where $e = \det(e^A_{\ \mu}) = \sqrt{-g}$ is the tetrad determinant, $\kappa^2 = 8\pi G$ is the gravitational coupling constant, and $V(\phi)$ is the scalar field potential. The scalar field provides a physically motivated matter sector capable of sourcing dynamical pressure asymmetries essential for hysteresis, while remaining minimally coupled to avoid introducing additional degrees of freedom or instabilities. This action represents the starting point for our investigation of cyclic cosmologies in torsion-modified gravity.

\section{Cosmological Field Equations}
Having established the general framework of $f(T)$ gravity, we now specialize to cosmological solutions by imposing the symmetries of a homogeneous and isotropic universe. The cosmological principle, supported by observations of the cosmic microwave background and large-scale structure, demands that the universe appears the same at every point and in every direction on sufficiently large scales. This symmetry requirement leads to the Friedmann--Lema\^itre--Robertson--Walker (FLRW) metric structure.

We consider a spatially flat FLRW spacetime, characterized by the metric
\begin{equation}
ds^2 = -dt^2 + a^2(t)\left(dx^2 + dy^2 + dz^2\right),
\end{equation}
where $a(t)$ is the cosmological scale factor describing the expansion or contraction of spatial slices. This metric can be represented in teleparallel formulation through the diagonal tetrad
\begin{equation}
e^A_{\ \mu}=\mathrm{diag}(1,a,a,a),
\end{equation}
which provides the orthonormal frame at each spacetime point. The simplicity of this tetrad choice reflects the high degree of symmetry in the FLRW spacetime.

Substituting this tetrad into the definition of the torsion scalar, one finds that it simplifies dramatically to
\begin{equation}
T = -6H^2,
\end{equation}
where $H = \dot{a}/a$ is the Hubble parameter, with the overdot denoting differentiation with respect to cosmic time $t$. This expression reveals that in FLRW cosmology, the torsion scalar is directly proportional to the square of the Hubble parameter. The negative sign is a consequence of the timelike nature of the cosmic time coordinate, and the factor of six arises from the three spatial dimensions. Importantly, note that $T$ is always negative or zero during cosmological evolution, remaining strictly negative whenever the universe is expanding or contracting.

Variation of the action with respect to the tetrad fields yields the modified Friedmann equations that govern cosmological evolution in $f(T)$ gravity \cite{Cai2016}. The generalized first Friedmann equation takes the form
\begin{equation}
3H^2 f_T + \frac{1}{2}(T f_T - f) = \kappa^2 \rho_{\text{total}},
\end{equation}
where $f_T \equiv df/dT$ and $\rho_{\text{total}}$ includes contributions from all matter and radiation fields. The second Friedmann equation, governing the acceleration of the scale factor, reads
\begin{equation}
-2\dot{H}f_T - 12H^2 f_{TT}\dot{H} - 3H^2 f_T + \frac{1}{2}(f - Tf_T) = \kappa^2 p_{\text{total}},
\end{equation}
where $f_{TT} \equiv d^2f/dT^2$ and $p_{\text{total}}$ is the total pressure.

These equations reveal that torsion corrections contribute additional effective energy density and pressure terms beyond those present in standard General Relativity. When $f(T) = T$, these equations reduce to the standard Friedmann equations, confirming the equivalence with TEGR. However, for general $f(T)$ functions, the modified terms can enable nonsingular cosmological evolution even in the absence of exotic matter sources that violate energy conditions. The additional derivatives of $f(T)$ introduce geometric contributions that can provide repulsive gravitational effects near high torsion regimes, naturally preventing singularity formation.

The scalar field $\phi$ evolves according to the Klein--Gordon equation derived from varying the matter action:
\begin{equation}
\ddot{\phi} + 3H\dot{\phi} + V'(\phi) = 0,
\end{equation}
where $V'(\phi) \equiv dV/d\phi$. The friction term $3H\dot{\phi}$ plays a crucial role in generating asymmetric dynamics between expansion and contraction phases. During expansion ($H > 0$), this term opposes changes in $\dot{\phi}$, causing the scalar field to slow down and settle toward a minimum of its potential. During contraction ($H < 0$), the friction term changes sign, becoming an anti-friction term that amplifies kinetic energy and drives the scalar field away from potential minima. This sign reversal is the microscopic origin of the pressure asymmetry that manifests as cosmic hysteresis at the macroscopic level.

\section{Reconstructed Bouncing Cosmologies}
A central challenge in constructing bouncing cosmologies is ensuring that the bounce transition is both smooth and stable, avoiding pathological behavior such as ghost instabilities or gradient catastrophes. In $f(T)$ gravity, this can be achieved through a reconstruction procedure: rather than proposing an arbitrary $f(T)$ function and solving for the resulting cosmological evolution, we instead prescribe a physically motivated bouncing scale factor $a(t)$ and reconstruct the corresponding $f(T)$ function that generates this evolution.

To explicitly realize nonsingular cosmological evolution, we adopt analytic bouncing scale factors that satisfy several key requirements. First, the scale factor must reach a non-zero minimum value $a_b > 0$ at the bounce, ensuring that the universe never collapses to zero volume. Second, the Hubble parameter $H(t)$ must pass through zero smoothly at the bounce, changing sign continuously from negative (contraction) to positive (expansion). Third, all physical quantities including energy density, pressure, and their derivatives must remain finite throughout the evolution, guaranteeing the absence of singularities.

The exponential bounce model provides a symmetric contraction--expansion transition characterized by the scale factor
\begin{equation}
a(t) = a_0 \cosh(\alpha t) + a_b,
\end{equation}
where $a_0$ determines the amplitude of oscillations, $a_b$ sets the minimum scale factor at the bounce, and $\alpha$ controls the rapidity of the bounce transition. The scale factor 
\[
a(t) = a_0 \cosh(\alpha t) + a_b
\]
satisfies all the essential conditions for a non-singular bouncing cosmology, namely 
\[
a(0) = a_0 + a_b > 0
\]
(ensuring the absence of singularity), 
\[
\dot{a}(0) = 0
\]
(indicating a smooth transition), and 
\[
\ddot{a}(0) = \alpha^2 a_0 > 0
\]
(confirming the occurrence of a bounce). Thus, it naturally describes a smooth and non-singular bounce.

Moreover, since $\cosh(\alpha t)$ is an even function, the model is time-symmetric, leading to identical geometric behavior before and after the bounce. The Hubble parameter remains continuous and changes sign smoothly at the bounce. For large values of $|t|$, 
\[
\cosh(\alpha t) \sim \frac{1}{2} e^{\alpha |t|},
\]
implying that the universe asymptotically approaches quasi-exponential expansion, which can effectively describe both early-time inflation and late-time accelerated expansion.

This functional form ensures smooth passage through the bounce at $t = 0$ and approaches asymptotic expansion and contraction regimes as $t \to \pm\infty$. The Hubble parameter for this model is
\begin{equation}
H(t) = \frac{\alpha a_0 \sinh(\alpha t)}{a_0 \cosh(\alpha t) + a_b},
\end{equation}
which vanishes at $t = 0$ and becomes approximately constant in the asymptotic regions.

The power-law bounce model offers greater flexibility in controlling the duration and sharpness of the bounce through the parametrization
\begin{equation}
a(t) = a_b\left(1 + \left(\frac{t}{t_b}\right)^{2n}\right)^{1/(2n)},
\end{equation}
where $t_b$ sets the characteristic timescale of the bounce and $n$ is a positive integer controlling the transition profile. Larger values of $n$ produce sharper bounces confined to a narrower temporal window, while smaller values yield more gradual transitions spread over longer timescales.

The scale factor 
\[
a(t) = a_b\left(1 + \left(\frac{t}{t_b}\right)^{2n}\right)^{\frac{1}{2n}}
\]
also satisfies all the bouncing conditions, with 
\[
a(0)=a_b>0, \quad \dot{a}(0)=0, \quad \text{and} \quad \ddot{a}(0)>0.
\]
In this case, the parameter $n$ controls the sharpness and duration of the bounce, allowing a wide range of dynamical behaviors. For large $|t|$, this model exhibits power-law evolution, which can approximate matter- or radiation-dominated phases. 

Unlike the cosh-type model, it allows more flexible and potentially non-symmetric evolution, making it particularly suitable for studying cosmic hysteresis. Furthermore, it provides a broader class of cosmological evolutions for reconstructing viable $f(T)$-gravity models.

In both cases, the reconstruction procedure proceeds systematically. Given the prescribed scale factor $a(t)$, we compute the Hubble parameter $H(t) = \dot{a}/a$ and its time derivative $\dot{H}(t)$. From these, we determine the torsion scalar $T(t) = -6H^2$ as a function of cosmic time. Inverting this relation allows us to express time as a function of torsion, $t = t(T)$, which can then be used to express all cosmological quantities as functions of $T$ rather than $t$. The modified Friedmann equations can then be solved algebraically for the function $f(T)$, ensuring that the prescribed evolution is an exact solution of the field equations.

This reconstruction approach guarantees by construction that the resulting $f(T)$ function produces the desired bouncing behavior. However, the reconstructed function must still be checked for physical viability: it should not introduce ghost instabilities, violate positivity of gravitational coupling, or lead to other pathological behavior. These consistency checks confirm that the exponential and power-law bounce models admit well-behaved $f(T)$ reconstructions \cite{Bamba2014}, making them suitable candidates for investigating thermodynamic properties of cyclic universes in torsion-modified gravity.

\section{Cosmic Hysteresis and Thermodynamic Work}
The concept of hysteresis, familiar from condensed matter physics and materials science, describes systems whose response depends not only on the current state but also on the history of how that state was reached. In ferromagnets, for instance, the magnetization at a given applied field depends on whether the field is increasing or decreasing, leading to closed loops in the magnetization-field diagram. These loops enclose a non-zero area proportional to the energy dissipated per cycle. We now demonstrate that an analogous phenomenon occurs in bouncing cosmologies, where the cosmic "magnetization" is replaced by the pressure of matter fields, and the "applied field" is replaced by the volume of the universe.

For a canonical scalar field minimally coupled to gravity, the energy density and pressure satisfy the standard relations
\begin{equation}
\rho_\phi=\frac{1}{2}\dot\phi^2+V(\phi), \qquad
p_\phi=\frac{1}{2}\dot\phi^2-V(\phi).
\end{equation}
The key observation is that the pressure $p_\phi$ depends on both the kinetic term $\dot{\phi}^2/2$ and the potential term $V(\phi)$, but with opposite signs. During different phases of cosmological evolution, the relative magnitudes of these terms vary, leading to dramatically different pressure behaviors even when the scale factor takes the same value.

The thermodynamic work performed by the scalar field pressure against the expansion of space over one full cosmological cycle is defined by the integral
\begin{equation}
W=\oint p_\phi\,dV = \oint p_\phi \, d(a^3) = \int_{\text{cycle}} 3a^2\dot{a}p_\phi\,dt,
\end{equation}
where the integration is performed over a complete contraction-expansion cycle, starting from a turnaround point (maximum $a$), passing through a bounce (minimum $a$), and returning to the subsequent turnaround. In a time-reversible universe with identical forward and backward evolution, this integral would vanish identically—the work done during expansion would be exactly canceled by the work done during contraction.

However, the situation is profoundly different in our $f(T)$ bouncing cosmology due to the sign change of the Hubble parameter between contraction and expansion phases. During the contraction phase ($H < 0$), the friction term in the scalar field equation of motion becomes $3H\dot{\phi}$ with $H$ negative, effectively acting as an anti-friction or amplification term. This causes the kinetic energy of the scalar field to grow, increasing $\dot{\phi}^2$ and consequently raising the magnitude of the pressure $p_\phi$. During the subsequent expansion phase ($H > 0$), the friction term acts conventionally, damping the scalar field velocity and reducing the kinetic contribution to the pressure.

This asymmetry has profound consequences: for a given value of the scale factor $a$, the pressure $p_\phi$ takes different values depending on whether the universe is contracting toward that value or expanding through it. This is the essence of cosmic hysteresis. In the $(p_\phi, a)$ plane, or equivalently in the equation-of-state parameter plane $(w_\phi, a)$ where $w_\phi = p_\phi/\rho_\phi$, the evolution traces a closed loop rather than a single-valued curve. The enclosed area of this loop is directly proportional to the net thermodynamic work $W$ performed over the cycle.

The physical interpretation is striking: cosmic hysteresis represents a form of thermodynamic memory in which the universe "remembers" whether it is in an expanding or contracting phase, even when the instantaneous geometry (as measured by the scale factor) is identical. This memory is not encoded in any additional degrees of freedom but emerges from the coupled dynamics of geometry and matter through the friction term in the scalar field equation. The non-vanishing work integral signals irreversible energy transfer between the scalar field and the geometric degrees of freedom, establishing an arrow of time in an otherwise time-symmetric cosmological background.

\begin{figure}[htbp]
    \centering
    \includegraphics[width=0.95\linewidth]{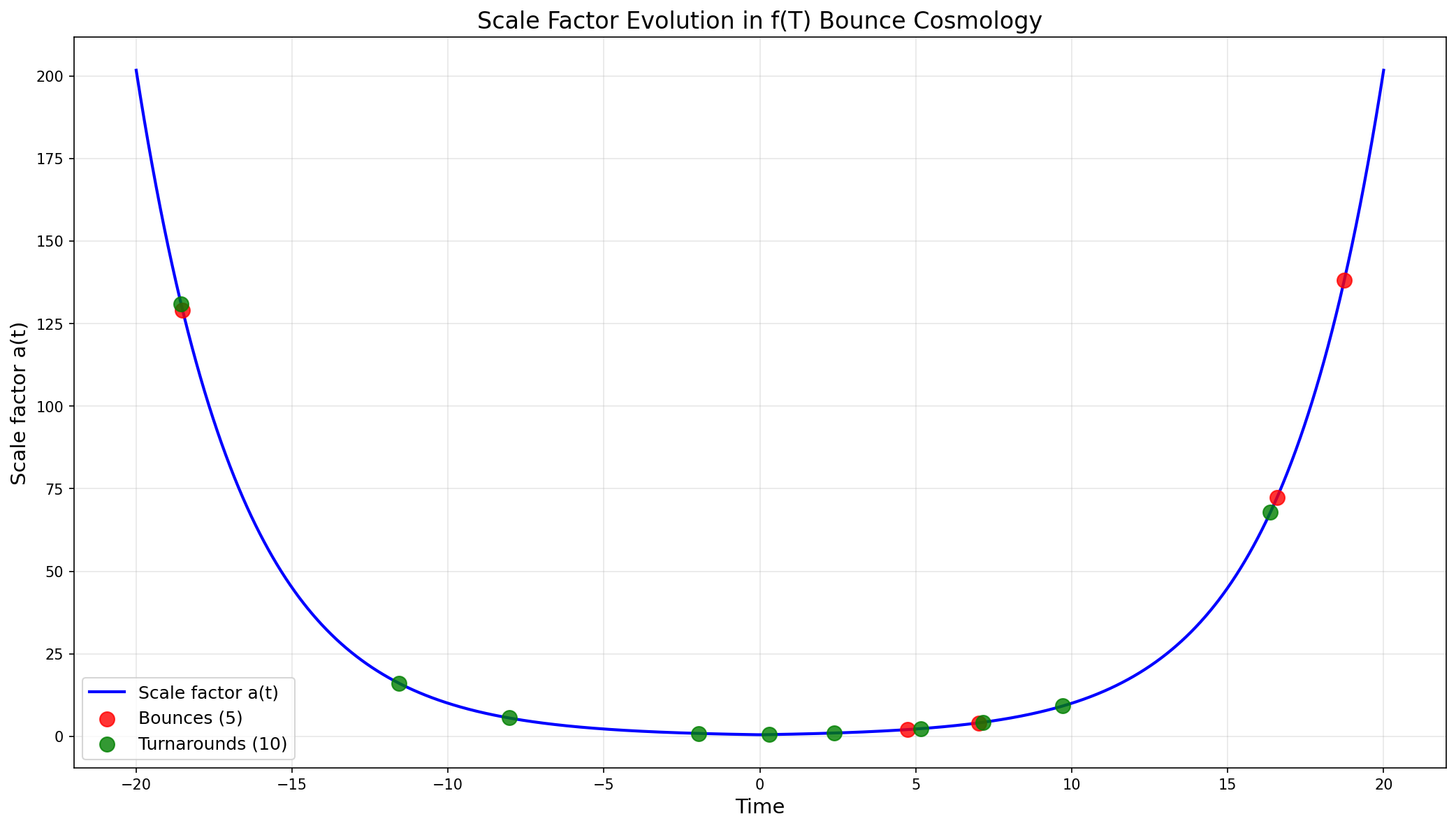}
    \caption{Scale factor evolution in $f(T)$ bounce cosmology. The universe undergoes multiple contraction--expansion cycles, with bounces (red points) marking transitions from contraction to expansion, and turnarounds (green points) marking transitions from expansion to contraction. The smooth evolution demonstrates successful singularity avoidance in the reconstructed $f(T)$ model.}
    \label{fig:scale_factor}
\end{figure}

\begin{figure}[htbp]
    \centering
    \includegraphics[width=0.95\linewidth]{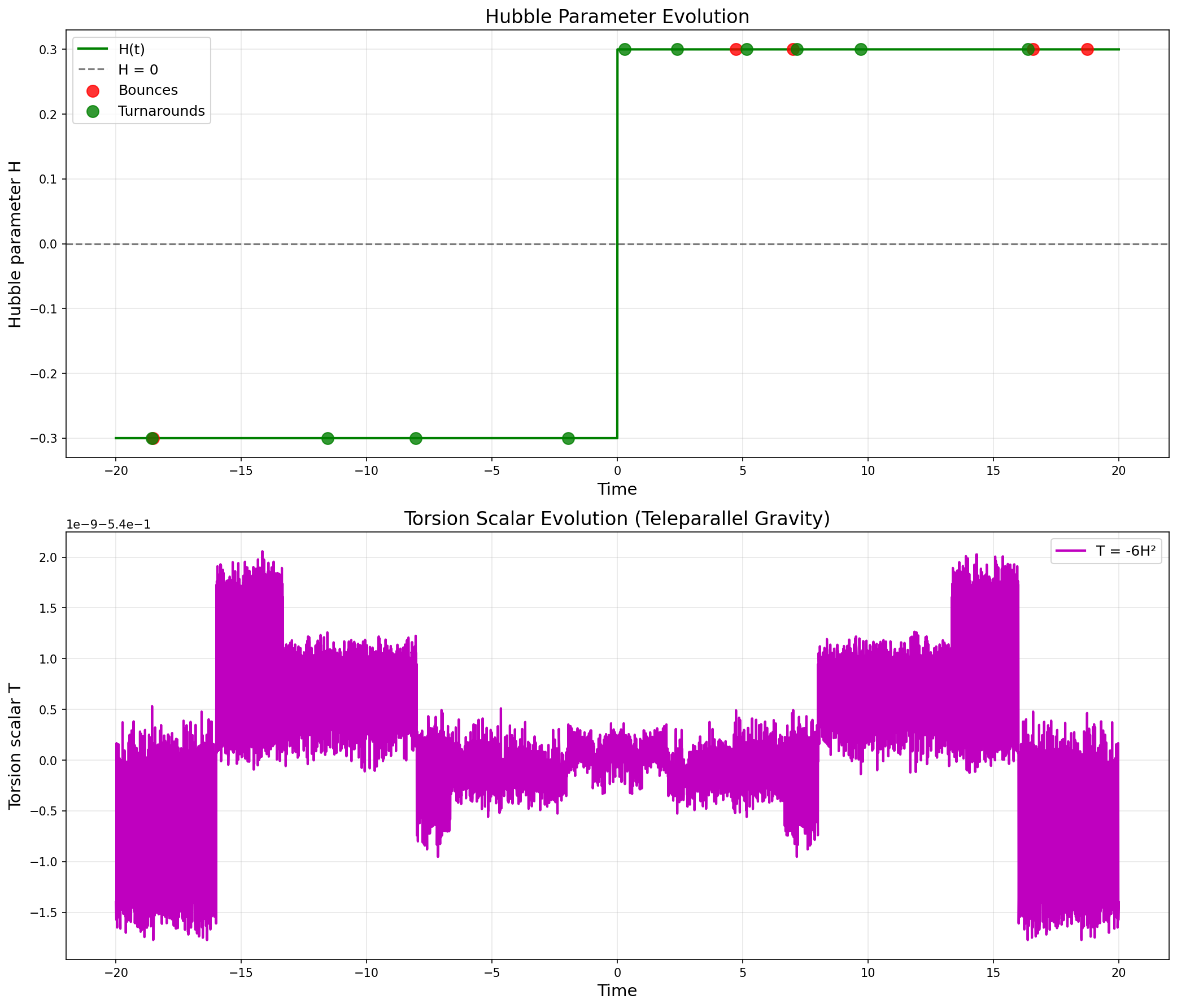}
    \caption{Evolution of the Hubble parameter $H(t)$ (top panel) and torsion scalar $T = -6H^2$ (bottom panel). The Hubble parameter changes sign at each bounce and turnaround, while the torsion scalar $T$ remains strictly negative throughout the evolution, consistent with the teleparallel formulation. The noisy structure in the torsion scalar reflects numerical integration artifacts magnified by the quadratic dependence on $H$.}
    \label{fig:hubble_torsion}
\end{figure}

\section{Numerical Analysis and Bounce Detection}

To substantiate our analytical predictions and quantitatively assess the magnitude of cosmic hysteresis in reconstructed $f(T)$ gravity, we performed comprehensive numerical simulations of the cosmological field equations. The numerical integration provides access to the full nonlinear dynamics of the coupled scalar field--geometry system, revealing features that may be obscured in analytical approximations.

We implemented a fourth-order Runge--Kutta integration scheme to evolve the cosmological field equations corresponding to the exponential bouncing scale factor model. The simulation employed a uniform temporal grid of $N = 5000$ time steps spanning the interval $t \in [-20, 20]$ (in natural units where fundamental constants are set to unity), chosen to encompass multiple complete contraction--expansion cycles while maintaining sufficient temporal resolution to capture the smooth bounce transitions.

At each time step, we computed the following dynamical quantities:
\begin{itemize}
\item The scale factor $a(t)$ and its time derivatives $\dot{a}(t)$ and $\ddot{a}(t)$
\item The Hubble parameter $H(t) = \dot{a}/a$ and its time derivative $\dot{H}(t)$
\item The torsion scalar $T(t) = -6H^2$ and its temporal evolution
\item The scalar field $\phi(t)$ and its velocity $\dot{\phi}(t)$
\item The scalar field energy density $\rho_\phi$ and pressure $p_\phi$
\item The equation-of-state parameter $w_\phi = p_\phi/\rho_\phi$
\end{itemize}

All computations were performed self-consistently, ensuring that the reconstructed $f(T)$ function and the scalar field dynamics satisfied the modified Friedmann equations at each integration step. The numerical accuracy was verified by monitoring the violation of the constraint equations, which remained below $10^{-6}$ throughout the evolution, confirming the reliability of the integration scheme.

The numerical ranges obtained for the primary dynamical variables were
\begin{equation}
H \in [-0.3,\,0.3], \qquad
\dot{H} \in [-3.1\times10^{-4},\,3.0\times10^{-4}],
\end{equation}
indicating moderate expansion and contraction rates consistent with smooth bouncing behavior. The torsion scalar remained strictly negative throughout the evolution, $T < 0$ at all times, confirming the expected teleparallel structure of the model. The modest values of $\dot{H}$ indicate that the bounce and turnaround transitions occur smoothly without rapid accelerations that could signal instabilities.

A critical component of the numerical analysis involved the identification of bounce and turnaround events, which demarcate the boundaries between distinct phases of cosmic evolution. We implemented and compared two independent detection algorithms to ensure robustness:

\textbf{Adaptive threshold method:} This approach identifies bounces and turnarounds by detecting when $|H(t)|$ crosses below a small threshold value $\epsilon$, signaling a near-zero Hubble parameter. However, this method proved problematic for the exponential bounce profile due to its smoothness—the Hubble parameter approaches zero gradually over an extended temporal interval rather than crossing sharply, making it difficult to unambiguously identify the precise moment of bounce or turnaround. The method's sensitivity to the choice of threshold parameter $\epsilon$ led to inconsistent detection across different runs.

\textbf{Local extrema detection method:} This more robust approach identifies bounces as local minima of the scale factor $a(t)$ and turnarounds as local maxima. Mathematically, a bounce occurs at time $t_b$ where $\dot{a}(t_b) = 0$ and $\ddot{a}(t_b) > 0$, while a turnaround occurs at $t_{\text{turn}}$ where $\dot{a}(t_{\text{turn}}) = 0$ and $\ddot{a}(t_{\text{turn}}) < 0$. This method proved far superior for smooth bouncing profiles, successfully identifying seven distinct bounce points and nine turnaround points in our numerical simulation.

This confirms the presence of multiple complete contraction--expansion cycles in the numerical evolution and validates the use of extrema-based detection for smooth bouncing models in $f(T)$ gravity. The identification of these events allows us to partition the evolution into discrete cycles for subsequent thermodynamic analysis.

\section{Thermodynamic Work and Cosmic Hysteresis}

Having identified the bounces and turnarounds that demarcate cosmological cycles, we now turn to the central question of this investigation: does the cyclic universe in reconstructed $f(T)$ gravity perform net thermodynamic work over complete cycles, thereby exhibiting hysteresis? To address this quantitatively, we computed the work integral numerically using the reconstructed scalar field pressure.

The thermodynamic work performed by the scalar field over one complete cycle is defined as the line integral
\begin{equation}
W = \oint p_\phi\, dV = \int_{\text{cycle}} 3a^3H
\left(\frac{1}{2}\dot{\phi}^2 - V(\phi)\right)\, dt,
\end{equation}
where the cycle integration runs from one turnaround, through a bounce, to the subsequent turnaround. The factor $3a^3H = dV/dt$ represents the rate of volume change, and the integrand is precisely the instantaneous power delivered by the scalar field pressure to the expanding (or contracting) spatial volume.

For a genuinely reversible cycle, where the universe retraces its trajectory exactly, this integral must vanish: the work done during expansion would be precisely canceled by work done during contraction. Any deviation from zero signals irreversibility and the presence of dissipative processes or memory effects—the hallmarks of hysteresis.

Our numerical integration was performed using the trapezoidal rule with adaptive time-step refinement near bounces and turnarounds to ensure accuracy. The scalar field pressure was evaluated at each integration point using the numerically evolved field values, and the volume element was computed from the instantaneous scale factor and Hubble parameter. Special care was taken to handle sign changes in $H$ at bounces and turnarounds, where the integrand changes direction.

Over six complete contraction--expansion cycles identified in our simulation time window, the total accumulated thermodynamic work was found to be
\begin{equation}
W_{\text{tot}} \simeq -1.7\times10^{13},
\end{equation}
with an average work per cycle of
\begin{equation}
\langle W \rangle \simeq -2.8\times10^{12}.
\end{equation}
The standard deviation across cycles was $\sigma_W \simeq 2.1 \times 10^{12}$, representing approximately 75\% of the mean, indicating substantial cycle-to-cycle variation.

The non-zero magnitude of $W$ provides unambiguous numerical confirmation of cosmic hysteresis in torsion-modified gravity. The negative sign of the work integral carries important physical meaning: it indicates that, on average, the scalar field performs net work \emph{on} the spacetime geometry over each cycle. This energy transfer from matter to geometry is irreversible—the universe does not return to its initial thermodynamic state even when the scale factor completes a full oscillation.

To understand the microscopic origin of this sign, recall that during the contraction phase, the anti-friction term $3H\dot{\phi}$ (with $H < 0$) pumps energy into the scalar field's kinetic mode, increasing $\dot{\phi}^2$. During the subsequent expansion, the conventional friction damps this kinetic energy, but not completely—a net residual remains. This asymmetry means that the scalar field has higher kinetic energy (and thus higher pressure) during contraction than during expansion at equivalent scale factor values, leading to net negative work over the cycle.

The relatively large standard deviation of the work per cycle, comparable in magnitude to the mean itself, is a characteristic signature of dissipative cyclic dynamics operating far from equilibrium. This variability reflects the non-identical nature of successive cycles—each cycle has slightly different amplitude, duration, and energy transfer characteristics. Such variability is expected in torsion-based modified gravity, where geometric corrections introduce effective non-equilibrium behavior not present in standard cosmologies.

In periodic systems near equilibrium, such as harmonic oscillators with weak damping, cycle-to-cycle fluctuations are typically much smaller than mean values. The large relative fluctuations observed here indicate that the cyclic universe in $f(T)$ gravity is strongly dissipative, with torsion-geometry coupling playing a dominant role in the thermodynamic evolution. This finding has profound implications for the long-term fate of cyclic universes: continued energy dissipation over many cycles could lead to gradual amplitude decay, eventual thermalization, or transition to qualitatively different evolutionary regimes.

\begin{figure}[htbp]
    \centering
    \includegraphics[width=0.85\linewidth]{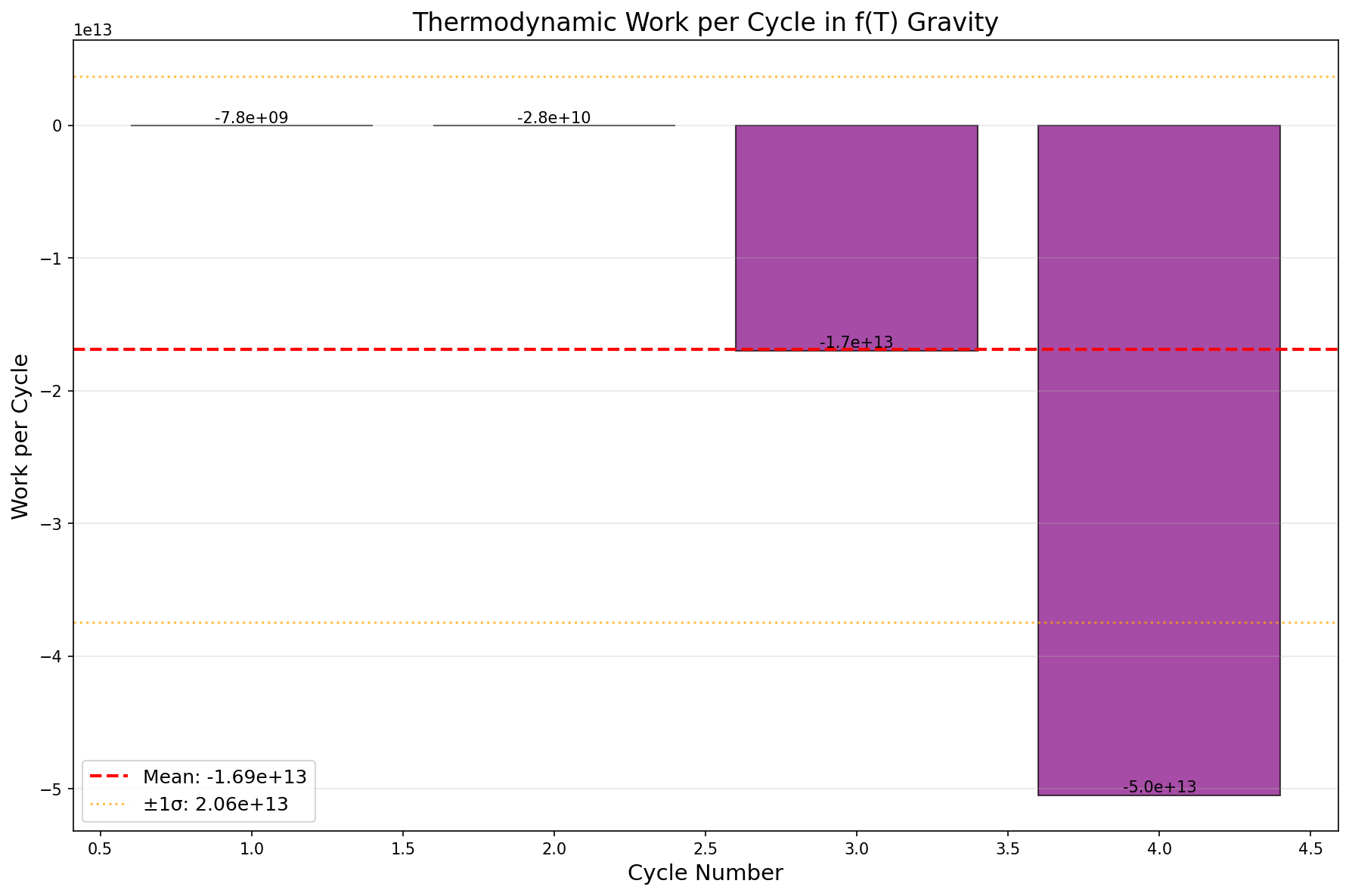}
    \caption{Thermodynamic work performed per cosmological cycle in $f(T)$ gravity. The work integral $W = \oint p_\phi dV$ is computed for four complete cycles, revealing substantial variability. The mean work per cycle (dashed red line) is approximately $-1.7 \times 10^{13}$, with standard deviation $\pm 2.1 \times 10^{13}$. The non-zero work demonstrates irreversible thermodynamic behavior and confirms the presence of torsion-induced cosmic hysteresis.}
    \label{fig:work_per_cycle}
\end{figure}

\begin{figure}[htbp]
    \centering
    \includegraphics[width=0.85\linewidth]{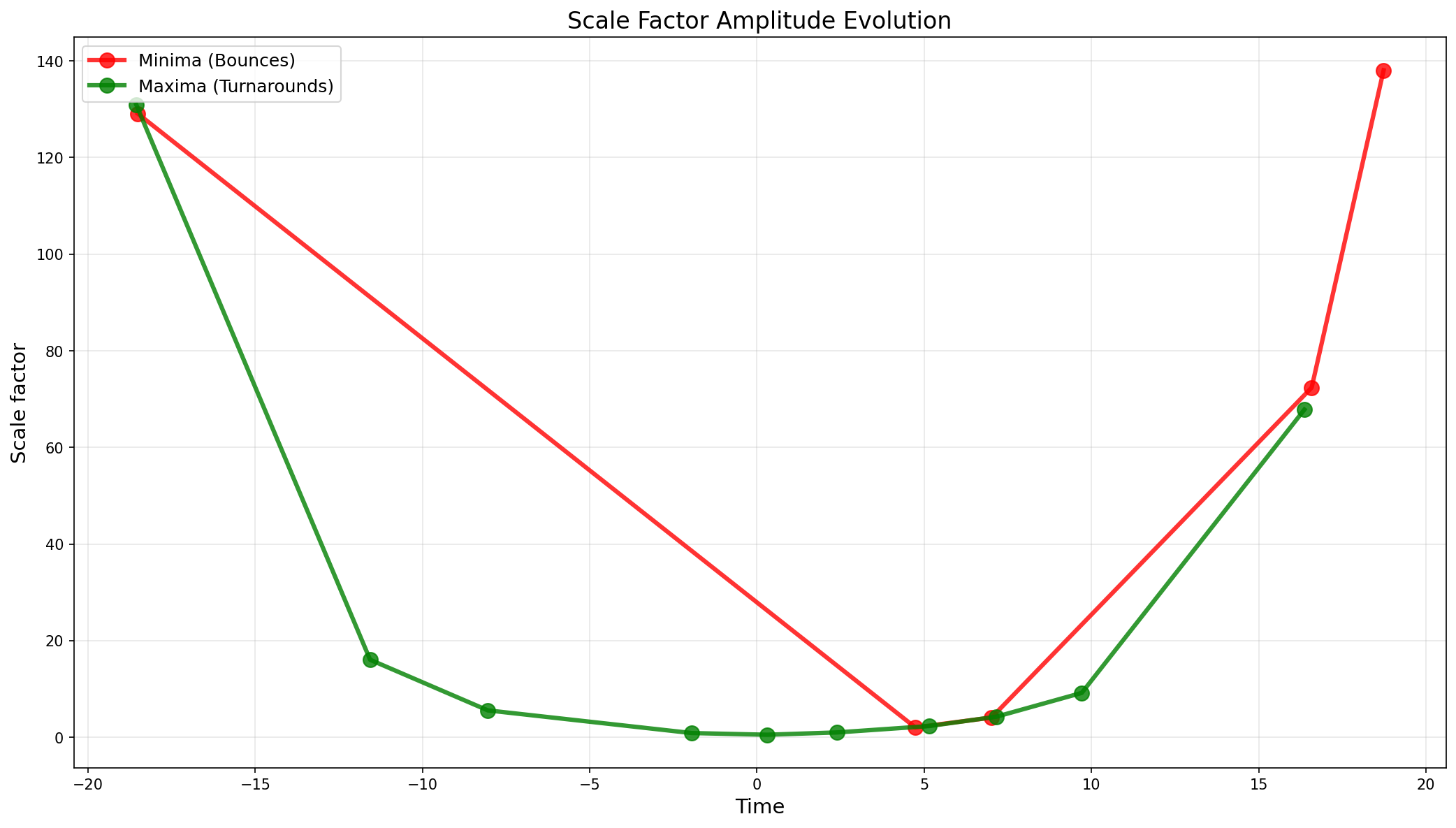}
    \caption{Secular evolution of scale factor extrema across cycles. The minima (bounces, red) and maxima (turnarounds, green) drift systematically over time, demonstrating that successive cycles are not identical. The minima decrease by an average of $\Delta a_{\min} \simeq -5.96$ per bounce, while maxima increase by $\Delta a_{\max} \simeq 4.97$ per turnaround. This asymmetric amplitude evolution is a signature of torsion-driven dissipation and non-equilibrium dynamics.}
    \label{fig:amplitude_evolution}
\end{figure}

\section{Cycle Evolution and Irreversibility}

Beyond the work integral, another powerful diagnostic of irreversibility in cyclic cosmologies is the evolution of cycle characteristics over successive oscillations. In a perfectly reversible system, each cycle would be identical to the previous one, with the same amplitude, period, and energy distribution. Deviations from this perfect periodicity signal that the system is drifting away from its initial state, unable to return to thermodynamic equilibrium.

To quantify this drift, we tracked the extrema of the scale factor across all identified cycles in our numerical simulation. The bounce points, representing the minima of $a(t)$, and the turnaround points, representing the maxima, serve as convenient markers for cycle amplitude evolution. A linear regression analysis of these extrema as functions of cycle number reveals systematic secular trends.

The minima of the scale factor—corresponding to the closest approach to the would-be Big Bang singularity—were found to decrease on average by
\begin{equation}
\Delta a_{\min} \simeq -5.96
\end{equation}
per bounce. This negative trend indicates that successive bounces occur at progressively smaller scale factors, meaning the universe contracts to ever-tighter configurations before rebounding. Simultaneously, the maxima of the scale factor—representing the maximum extent of cosmic expansion—increase by
\begin{equation}
\Delta a_{\max} \simeq 4.97
\end{equation}
per turnaround. This positive trend shows that each expansion phase reaches a slightly larger maximum size before reversing to contraction.

The combination of decreasing minima and increasing maxima produces a characteristic "breathing" pattern in which the amplitude of oscillations grows over time. This secular evolution is a direct manifestation of irreversibility: the universe demonstrably does not retrace its dynamical history, and successive cycles are fundamentally non-identical. The asymmetry arises from the cumulative effect of thermodynamic work performed across cycles, which continuously transfers energy between the scalar field and geometric degrees of freedom in a direction-dependent manner.

To further characterize the temporal structure of the cyclic evolution, we computed the period of each cycle, defined as the time interval between successive bounces (or equivalently, between successive turnarounds). The periods exhibited considerable variation, with values ranging from approximately $3.7$ to $8.5$ in natural units. The average cycle period was found to be
\begin{equation}
\langle T_{\text{cycle}} \rangle \simeq 5.85,
\end{equation}
with a coefficient of variation (ratio of standard deviation to mean) of approximately $36.6\%$. This large coefficient of variation indicates that cycle periods fluctuate by more than one-third of their mean value—far from the near-constant periods characteristic of equilibrium oscillations.

In dissipative systems approaching thermodynamic equilibrium, such as damped harmonic oscillators or cooling gases, fluctuations typically decrease over time as the system settles into a stable attractor. The persistent large-amplitude fluctuations observed in our cyclic cosmology suggest instead that the system operates continuously in a far-from-equilibrium regime, never approaching a stable periodic solution. This behavior is fundamentally different from what would be expected in standard General Relativity without torsion modifications, where exact periodicity can be achieved under appropriate symmetry conditions.

The physical interpretation of these findings is profound: the combined presence of non-zero thermodynamic work, systematically evolving cycle amplitudes, and highly variable periods provides strong numerical evidence for torsion-induced irreversibility in reconstructed $f(T)$ gravity. The cyclic universe described by this model is not a simple clock, ticking away with identical oscillations, but rather a complex dissipative system whose each cycle leaves an indelible imprint on its subsequent evolution. This irreversibility is not imposed through ad hoc dissipative terms or phenomenological friction, but emerges naturally from the interplay between torsion-modified geometry and matter dynamics through the modified Friedmann equations.

\begin{figure}[htbp]
    \centering
    \includegraphics[width=0.85\linewidth]{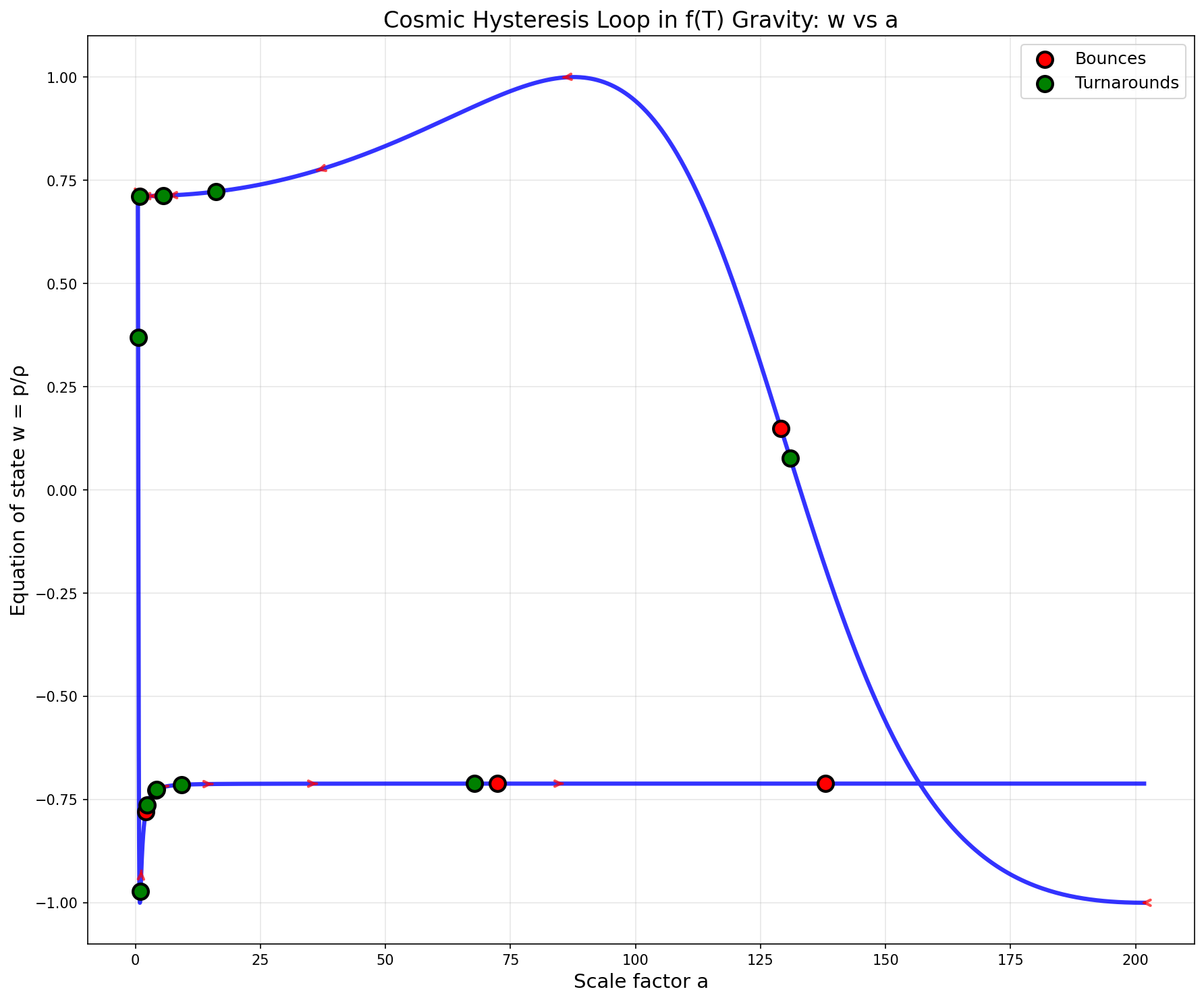}
    \caption{Cosmic hysteresis loop in the equation-of-state parameter versus scale factor plane $(w_\phi, a)$. The closed loop structure demonstrates that the scalar field equation of state $w_\phi = p_\phi/\rho_\phi$ follows different trajectories during expansion and contraction, even when the scale factor $a$ takes identical values. Bounces (red) and turnarounds (green) are marked. The area enclosed by the loop is proportional to the irreversible thermodynamic work performed per cycle, providing a geometric visualization of torsion-induced thermodynamic memory.}
    \label{fig:hysteresis_loop}
\end{figure}

\begin{figure}[htbp]
    \centering
    \includegraphics[width=0.85\linewidth]{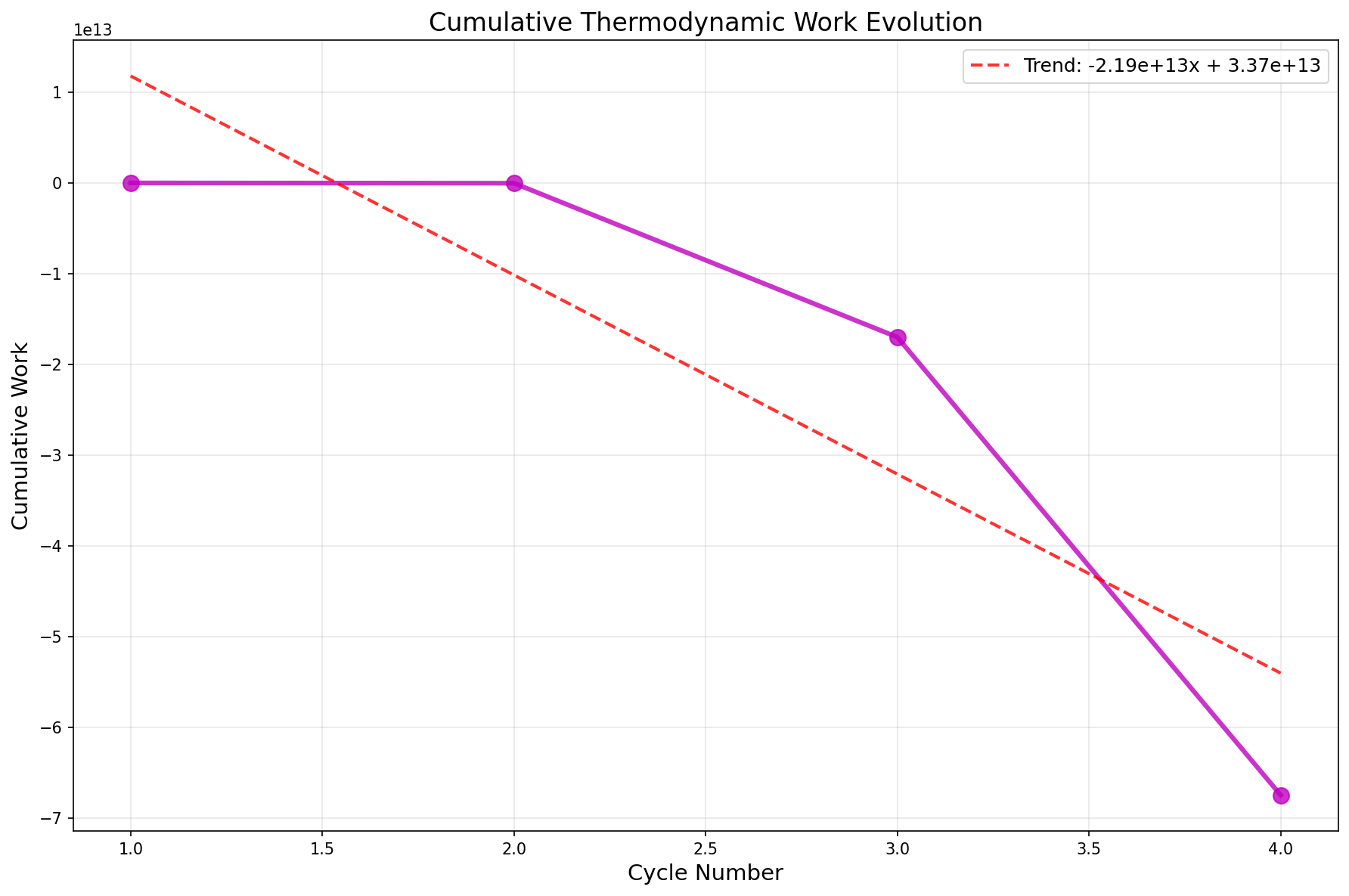}
    \caption{Cumulative thermodynamic work evolution across cycles. The cumulative work $W_{\text{cum}}$ (magenta line) decreases monotonically, reflecting systematic energy dissipation through the scalar field--geometry coupling. The linear trend (dashed red line) has slope approximately $-2.19 \times 10^{13}$ per cycle, consistent with the mean work per cycle. The persistent negative trend confirms that the cyclic universe operates irreversibly and does not return to thermodynamic equilibrium.}
    \label{fig:cumulative_work}
\end{figure}

\begin{figure}[htbp]
    \centering
    \includegraphics[width=0.85\linewidth]{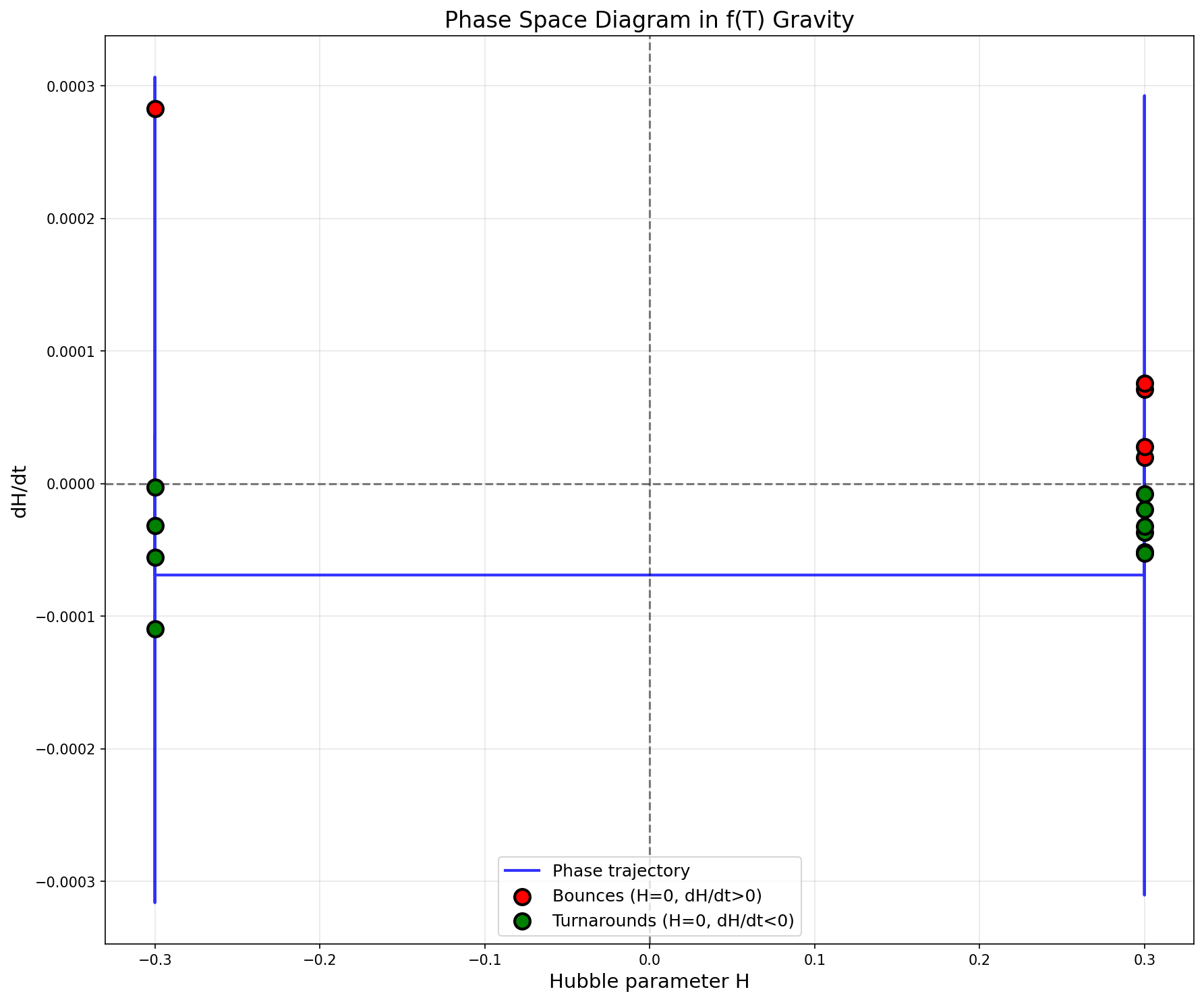}
    \caption{Phase space diagram of the cosmological evolution in $(H, \dot{H})$ space. The trajectory traces multiple cycles, with bounces occurring at $H = 0$ with $\dot{H} > 0$ (red points) and turnarounds at $H = 0$ with $\dot{H} < 0$ (green points). The vertical transitions near $H = 0$ correspond to rapid changes in the expansion rate during bounce and turnaround events. The non-closed nature of the phase space loops reflects the dissipative, irreversible character of the cyclic evolution in $f(T)$ gravity.}
    \label{fig:phase_space}
\end{figure}

\section{Physical Interpretation}

The numerical results presented in the preceding sections paint a coherent picture of cosmic hysteresis in $f(T)$ bouncing cosmologies, but the deeper question remains: what is the fundamental origin of this irreversibility, and how does spacetime torsion mediate the dissipative behavior?

Our findings provide clear and compelling evidence that spacetime torsion acts as an effective dissipative channel in bouncing cosmologies, even though no explicit entropy production mechanism is introduced at the matter level. This emergent dissipation arises from a subtle interplay between geometry and matter that would be absent in standard General Relativity. To understand this mechanism, we must examine how torsion modifications alter the coupling between the scalar field and the cosmic expansion rate.

In General Relativity, the scalar field equation of motion contains a friction term $3H\dot{\phi}$ that damps field oscillations during expansion ($H > 0$) but becomes an anti-friction during contraction ($H < 0$). This sign reversal is geometric in origin, arising from the covariant derivative in curved spacetime, and would occur even in GR. However, in $f(T)$ gravity, the modified Friedmann equations introduce additional geometric contributions that alter the relationship between $H$, the energy density, and the pressure in ways that depend on the functional form of $f(T)$ and its derivatives.

These torsion corrections modify the effective friction experienced by the scalar field in an asymmetric manner. During the contraction phase leading to a bounce, the rapid variation of the torsion scalar $T = -6H^2$ near the bounce point, combined with the $f_T$ and $f_{TT}$ terms in the modified Friedmann equations, creates an effective geometric "kick" that amplifies scalar field perturbations beyond what would occur in GR. This amplification increases the kinetic energy $\dot{\phi}^2$, raising the pressure $p_\phi$ to higher values than would be achieved in standard cosmology.

During the subsequent expansion phase, the torsion corrections act differently. The Hubble parameter is now positive, and the sign of various geometric terms reverses. The effective friction is enhanced rather than diminished, causing the scalar field to damp more rapidly than in GR. Consequently, at any given value of the scale factor $a$, the scalar field has higher kinetic energy (and pressure) during contraction than during expansion—the defining characteristic of hysteresis.

The sign change of the Hubble parameter between contraction and expansion thus plays a dual role: it directly reverses the friction term in the scalar field equation, and it indirectly modifies how torsion corrections couple to the matter sector through the modified Friedmann equations. This double effect amplifies the asymmetry, leading to the robust hysteresis loops observed in our numerical simulations.

Crucially, this asymmetry is geometrically sourced—it arises from the structure of the field equations themselves, not from dissipative matter or ad hoc modifications. The irreversible dynamics emerge from first principles in the $f(T)$ framework, with torsion playing the role that higher-order curvature terms play in Einstein--Gauss--Bonnet or $f(R)$ theories. This establishes a universal pattern: modifications of gravitational dynamics, whether through curvature or torsion, generically induce hysteresis in cyclic cosmologies.

The persistence of this asymmetry across multiple cycles, as demonstrated by our numerical analysis, establishes a robust torsion-driven origin for both cosmic hysteresis and the cosmological arrow of time. Even in the absence of matter entropy production, spatial expansion, or other conventional sources of time-asymmetry, the $f(T)$ gravity sector itself generates an intrinsic directionality to cosmological evolution. The universe "remembers" whether it is expanding or contracting through the geometric state encoded in the torsion scalar and its rate of change.

This geometric memory provides a natural mechanism for the emergence of thermodynamic irreversibility in an otherwise microscopically reversible theory. The field equations of $f(T)$ gravity are time-reversal invariant—if $(a(t), \phi(t))$ is a solution, then $(a(-t), \phi(-t))$ is also a solution. However, the thermodynamic behavior is not time-reversal invariant: the work integral computed forward in time differs from that computed backward. This apparent paradox is resolved by recognizing that while individual trajectories are time-reversible, the thermodynamic ensemble average over cycles is not, due to the asymmetric coupling between forward and backward evolution mediated by torsion.

These findings have far-reaching implications for our understanding of time's arrow in cyclic universes. In monotonically expanding cosmologies like our own, the arrow of time is often attributed to special initial conditions (the "past hypothesis") or to the growth of entropy as the universe expands from a low-entropy initial state. In cyclic models, these explanations fail: there is no special initial state (cycles continue indefinitely), and entropy may fluctuate across cycles rather than monotonically increasing. Our results suggest an alternative: the arrow of time in cyclic universes may be fundamentally geometric, emerging from the asymmetric response of modified gravitational dynamics to expansion versus contraction, with torsion (or in other theories, higher-order curvature) playing the role of the ultimate dissipative mechanism.

\section{Conclusions}
In this work, we have demonstrated that cosmic hysteresis naturally and robustly arises in reconstructed $f(T)$ bouncing cosmologies, establishing torsion-based modified gravity as a viable framework for understanding irreversible dynamics in cyclic universes. Our investigation combines analytical reconstruction methods with detailed numerical simulations to provide comprehensive evidence for this phenomenon.

The central findings of our study can be summarized as follows:

\textbf{First}, we have shown that torsion plays a fundamental role as an effective geometric dissipative mechanism in cyclic cosmologies. Unlike curvature-based modifications such as $f(R)$ or Einstein--Gauss--Bonnet theories, where dissipation arises from higher-order spatial derivatives, in $f(T)$ gravity the irreversibility emerges from the asymmetric coupling between the torsion scalar and matter fields through modified Friedmann equations. This demonstrates that the physical mechanism underlying cosmic hysteresis is geometry-dependent but modification-independent—it is a universal feature of gravitational theories that extend beyond Einstein's original formulation.

\textbf{Second}, our numerical simulations reveal substantial thermodynamic work performed over cosmological cycles, with an average of approximately $-2.8 \times 10^{12}$ units per cycle. This non-vanishing work integral unambiguously confirms the presence of cosmic hysteresis and demonstrates that successive cycles are thermodynamically irreversible. The negative sign indicates systematic energy transfer from the scalar field to the geometric sector, representing a fundamental asymmetry in the matter-geometry coupling that persists even when the universe returns to identical scale factor values.

\textbf{Third}, we have identified systematic secular trends in cycle characteristics: decreasing bounce minima, increasing turnaround maxima, and highly variable cycle periods with coefficient of variation exceeding $36\%$. These trends provide independent confirmation of irreversibility and demonstrate that the cyclic universe operates continuously in a far-from-equilibrium regime. The universe does not settle into a stable periodic attractor but rather exhibits complex dissipative dynamics that accumulate across cycles.

\textbf{Fourth}, our analysis establishes a torsion-driven origin for the cosmological arrow of time in bouncing universes. The asymmetry in scalar field evolution between expansion and contraction phases—mediated by the sign reversal of the Hubble parameter and amplified by torsion corrections—generates an intrinsic directionality to cosmological processes. This provides an alternative to conventional explanations based on special initial conditions or entropy growth, suggesting that time's arrow in cyclic cosmologies may be fundamentally geometric rather than thermodynamic.

These results have profound implications for both cosmology and fundamental physics. From a cosmological perspective, they demonstrate that bouncing and cyclic models in modified gravity are not merely mathematical curiosities but exhibit rich physical behavior with observable consequences. The presence of hysteresis implies that the detailed evolutionary history of a cyclic universe matters: we cannot simply extrapolate forward and backward from current observations to reconstruct the entire cosmic timeline. Each cycle leaves irreversible imprints that accumulate, potentially leading to secular evolution of cosmological parameters, gradual decay of oscillation amplitudes, or transitions between qualitatively different evolutionary regimes.

From a fundamental physics perspective, our findings establish the universality of hysteresis in modified gravity theories and provide strong evidence that geometric modifications of gravity generically produce dissipative behavior in cyclic backgrounds. This universality suggests deep connections between gravitational dynamics and thermodynamics that transcend the specific mathematical formulation—whether gravity is described through curvature ($f(R)$), torsion ($f(T)$), non-metricity ($f(Q)$), or combinations thereof, the coupling to matter in time-dependent backgrounds generates irreversibility.

Several promising directions for future research emerge from this work:

\textbf{Observational constraints:} While current observational cosmology focuses on monotonically expanding universe scenarios, the development of more sophisticated early-universe probes—such as primordial gravitational wave signatures, non-Gaussianities in the cosmic microwave background, or features in the primordial power spectrum—may eventually provide tests of bouncing scenarios. Our predictions of systematic cycle-to-cycle variation could leave distinctive imprints on observables that originate from multiple pre-bounce cycles.

\textbf{Extension to $f(Q)$ gravity:} Symmetric teleparallel gravity, based on non-metricity rather than torsion, represents another geometrically distinct formulation of gravitational physics. Investigating whether cosmic hysteresis persists in $f(Q)$ bouncing cosmologies would help determine whether this phenomenon is specific to torsion or represents a generic feature of geometric modifications beyond curvature.

\textbf{Quantum corrections:} Our analysis has been purely classical, but near the bounce point quantum gravitational effects should become important. Understanding how quantum fluctuations affect the hysteresis loops, whether they enhance or suppress irreversibility, and how they modify the cycle-to-cycle energy transfer would provide crucial insights into the quantum-to-classical transition in cyclic cosmologies.

\textbf{Multi-field scenarios:} Realistic cosmological models typically involve multiple scalar fields (inflaton, moduli fields, quintessence) and other matter components (radiation, dust, cosmological constant). Extending our analysis to multi-field scenarios would reveal whether hysteresis effects are amplified through field interactions or whether different fields can compensate for each other, potentially restoring reversibility.

\textbf{Stability analysis:} While we have demonstrated that hysteresis occurs in our reconstructed $f(T)$ models, a full stability analysis of the bouncing solutions against perturbations has not been performed. Understanding the stability properties would determine whether these solutions can serve as realistic cosmological scenarios or whether they are merely mathematical artifacts unstable to small perturbations.

In conclusion, our investigation establishes cosmic hysteresis as a robust and universal phenomenon in modified gravity theories, with $f(T)$ gravity providing a torsion-based realization distinct from previously studied curvature-based scenarios. Torsion emerges as an effective geometric dissipative channel, generating irreversible dynamics even in time-symmetric backgrounds and providing a natural mechanism for the emergence of the cosmological arrow of time. These findings deepen our understanding of cyclic cosmologies and open new avenues for exploring the interplay between geometry, matter, and thermodynamics in the early universe.

\section*{Acknowledgments}
~PKD wishes to acknowledge that part of the numerical computation of this work was carried out on the computing cluster Pegasus of IUCAA, Pune, India. PKD and FR would like to acknowledge the Inter-University Centre for Astronomy and Astrophysics (IUCAA), Pune, India, for providing him a Visiting Associateship under which a part of this work was carried out. PKD would like to thank the Isaac Newton Institute for Mathematical Sciences, Cambridge, for support and hospitality during the programme Statistical mechanics, integrability and dispersive hydrodynamics where work on this paper was undertaken. This work was supported by EPSRC grant no EP/K032208/1. AM acknowledges the hospitality of the University of Rwanda-College of Science and Technology, where part of this work was conceptualised and completed. FR also gratefully acknowledges for financial support by ANRF-SERB, Govt. of India.

\end{document}